\documentclass[12pt]{article}

\pdfoutput=1

\usepackage{latexsym}
\usepackage{epsfig,amssymb,euscript}
\usepackage{amsmath,amsfonts}
\usepackage{mathrsfs}
\usepackage{ccaption}             
\usepackage[usenames]{color}

\usepackage{ dsfont }
\usepackage[
      colorlinks=true,
      linkcolor=blue,
      urlcolor=blue,
      filecolor=black,
      citecolor=red,
      pdfstartview=FitV,
      pdftitle={},
        pdfauthor={Simon A. Gentle, Benjamin Withers},
        pdfsubject={},
        pdfkeywords={},
        pdfpagemode=None,
        bookmarksopen=true
      ]{hyperref}

  \captionnamefont{\bfseries}
  \captiontitlefont{\small\sffamily}
  \captiondelim{: }
  \hangcaption

\def\be{\begin{equation}}
\def\ee{\end{equation}}
\def\bea{\begin{eqnarray}}
\def\eea{\end{eqnarray}}
\def\tr{\text{tr}}
\def\id{\mathds{1}}

\title{\bf Superconducting instabilities of R-charged black branes}

\author{\normalsize Simon A.~Gentle and Benjamin Withers\\ \\
        \small \it Centre for Particle Theory \& Department of Mathematical Sciences, \\
        \small \it Science Laboratories, South Road, Durham DH1 3LE, United Kingdom. \\ \\
        \normalsize\href{mailto:s.a.gentle@durham.ac.uk}{\texttt{s.a.gentle}}\texttt{, }\href{mailto:b.s.withers@durham.ac.uk}{\texttt{b.s.withers@durham.ac.uk}}}
\date{July 2012}

\begin{document}

\setlength{\baselineskip}{18pt}

\maketitle
\begin{picture}(0,0)(0,0)
\put(350,260){DCPT-12/29}
\end{picture}
\vspace{-36pt}
\thispagestyle{empty}                        

\begin{abstract}
We explore superconducting instabilities of black branes in $SO(6)$ gauged supergravity at finite temperature and finite R-charge densities.
We compute the critical temperatures for homogeneous neutral and superconducting instabilities in a truncation of 20 scalars and 15 gauge fields as a function of the chemical potentials conjugate to the three $U(1)$ charges in $SO(6)$. We find that despite the imbalance provided by multiple chemical potentials there is always at least one superconducting black brane branch, emerging at a temperature where the normal phase is locally thermodynamically stable. We emphasise that the three-equal charge solution, Reissner-Nordstr\"om,  is subdominant to a thermodynamically unstable black brane at sufficiently low temperatures --- a feature which is hidden in an equal charge truncation.
\end{abstract}

\pagebreak
\setcounter{page}{1}

%%%%%%%%%%%%%%%%%%%%%%%%%%%%%%%%%%%%%%%%%%%%%%
\section{Introduction}
\label{sec:Introduction}
%%%%%%%%%%%%%%%%%%%%%%%%%%%%%%%%%%%%%%%%%%%%%%

In the holographic approach to condensed matter physics there is great freedom in how a bulk theory is constructed. This has the advantage of allowing the construction of holographic models which capture a wide variety of physical phenomena. The drawback, of course, is that the detailed phenomenology of such constructions typically depends on the particular model employed. For this reason a desirable feature of any phenomenological holographic programme is the existence of some `universal result', at least for a particular class of bulk theories. A celebrated example is the ratio of shear viscosity to entropy density for strongly coupled, isotropic holographic plasmas arising from two derivative bulk Lagrangians \cite{Kovtun:2003wp, Buchel:2003tz}.

Holographic superconductivity on the other hand has no simple universal sector; in the minimal example of a single $U(1)$-charged bulk scalar field there is freedom to specify the potential and gauge field coupling. In a bottom-up approach, whilst the small-field expansion of these functions can be easily parameterised, the behaviour away from the critical temperature depends strongly on their nonlinear characteristics. Turning to a top-down approach one may hope to overcome such difficulties. However, as observed in \cite{Donos:2011ut}, different consistent sub-truncations of a particular top-down model exhibit radically different results for the behaviour of the superconducting order parameter with temperature. To understand if there are any generic features for the landscape of consistent holographic superconductors, one must delve deeper into larger families of consistent truncations. See \cite{Denef:2009tp, Donos:2011ut, Aprile:2011uq, DallAgata:2011aa} for some interesting work along these lines.

In this paper we take a step in this direction by considering a truncation of $\mathcal{N}=8$ $SO(6)$ gauged supergravity in 5D containing 20 scalars and 15 gauge fields. The truncation contains a three-charge family of black branes \cite{Behrndt:1998jd}, dual to a normal phase, about which we look for the emergence of new branches of superconductors. Such branches are indicated by the existence of marginal modes for the various charged scalar fields at finite temperature. This calculation can be viewed as a first step towards understanding the sequence of preferred phases of the full supergravity theory.

When cooled sufficiently the black branes we employ to model the normal phase become thermodynamically unstable.\footnote{At high temperatures in the regime of local thermodynamic stability there are indications that this system is only metastable \cite{Yamada:2006rx,Yamada:2008em,Hollowood:2008gp}. With the interpretation of the three R-charges as angular momenta of D3-branes spinning in the $S^5$ of the full 10D theory, this instability corresponds to the one-by-one fragmentation of D3-branes from the rotating stack. The superconducting branches we seek in this work are local instabilities of the black brane solutions, and whilst they may be subject to arguments vis-\`a-vis metastability, a study of the critical temperatures and chemical potential dependence is not directly affected by it. The family of black branes used here represents a convenient local minimum for our purposes.} Note that in many cases the stable black brane simply ceases to exist at low temperatures, and so the system is largely immune to any $T=0$ test of stability, such as Breitenlohner-Freedman (BF) bound violation for any near horizon $AdS_2\times \mathbb{R}^3$. We focus on superconducting instabilities that occur in the normal phase at sufficiently high temperatures where it is locally thermodynamically stable. We find that a superconducting black brane branch always exists in this region of the normal phase.

In addition, the black brane solutions we employ for the normal phase have three R-charges, which allows us to explore the dependence of the various superconducting phase transitions on the conjugate chemical potentials.  Some modes in the truncation are charged under a single $U(1)$ and do not couple to the other gauge fields at the linearised level. As such, turning on an additional chemical potential for another $U(1)$ provides an effective holographic description of a state with a density imbalance. An example in condensed matter physics is an imbalanced Fermi mixture in which there is a mismatch of chemical potentials for spin-up and spin-down constituents \cite{citeulike:2350785}. Phenomenological holographic models of such mixtures were investigated in bottom-up bulk and probe brane constructions in \cite{Erdmenger:2011hp, Bigazzi:2011ak}, where the additional $U(1)$ plays the role of up-down spin imbalance. One feature of these models is that the superconducting phase ceases to exist if the imbalance is too great.

Here we have an imbalance as a natural consequence of the three chemical potentials.  
As the imbalance for a given superconducting mode is increased, its critical temperature decreases until it reaches the threshold of thermodynamic instability, at finite temperature. However, increasing the imbalance for one mode decreases the imbalance for another, and is a consequence of the $SO(6)$ symmetry. Surprisingly there is no set of chemical potentials for which the imbalance is simultaneously too great for all modes. 

The paper proceeds as follows. In section \ref{sec:trunc} we present the consistent truncation and the three-chemical potential black brane solution we will be employing as the background normal phase. In section \ref{sec:fluctuations} we calculate the field equations for homogeneous fluctuations of the other fields in the truncation about this background that do not source gravitational fluctuations. The structure of the equations for the fluctuations and the instabilities found naturally leads to two different types of mode: modes that are continuously connected to charged modes are discussed in \ref{sec:minusmodes}, while those that are continuously connected to neutral modes are discussed in section \ref{sec:plusmodes}. In both cases we map out the regions in the space of chemical potentials which are unstable to each mode and detail their critical temperatures. We conclude in section \ref{sec:discussion}.

%%%%%%%%%%%%%%%%%%%%%%%%%%%%%%%%%%%%%%%%%%%%%%
\section{Truncation and the normal phase}\label{sec:trunc}
%%%%%%%%%%%%%%%%%%%%%%%%%%%%%%%%%%%%%%%%%%%%%%
In addition to gravity, the bosonic sector of $\mathcal{N}=8$ $SO(6)$ gauged supergravity in 5D \cite{Pernici:1985ju, Gunaydin:1984qu,Gunaydin:1985cu} contains 42 scalars, 15 $SO(6)$ gauge fields as well as twelve 2-form gauge potentials. The scalars are organised into irreps of $SO(6)$ which have masses $m^2L^2 = -4,-3,0$ in an $AdS_5$ vacuum, the lightest of which saturates the BF-bound. We will employ a further consistent  truncation of this theory down to 15 gauge fields and 20 of the BF-bound saturating scalars. In 10D ingredients this truncation corresponds to retaining only the metric and 5-form \cite{Cvetic:2000nc}. 

This truncation provides a convenient starting point in our search for superconducting black brane branches. First, the truncation contains a three-parameter family of black brane backgrounds which we may employ as the normal phase. Second, only the lightest scalars remain and these should display a greater susceptibility to forming condensates in the black brane backgrounds.\footnote{Superconducting branches for special cases of the other scalar field sectors are investigated in \cite{Aprile:2011uq}.} Finally, we are able to make contact with special cases studied elsewhere at the nonlinear level.
 
The 20 scalars and 15 gauge fields can be packaged as a symmetric unimodular $SO(6)$ tensor $T_{ij}$ and an antisymmetric $SO(6)$ tensor $A_{ij}$, respectively.  The equations of motion for this consistent truncation may be derived from the following action, with the adjoint $SO(6)$ indices written in matrix form:
\be \label{eq:action}
S = \frac{1}{16\pi G_5} \int d^5x \sqrt{-g} \left(R - \frac{1}{4} \tr\, T^{-1} \left(D_\mu T\right) \,T^{-1} \left(D^\mu T\right) +\frac{1}{8} \tr\, T^{-1} F_{\mu\nu} T^{-1} F^{\mu\nu} - V \right)
\ee
We have omitted a Chern-Simons term that will not be important for the electric, homogeneous superconductors sought here.  The scalar potential is given by
\be
V = \frac{g_c^2}{2}\left(2\tr T^2 - (\tr T)^2\right),
\ee
with the $SO(6)$ gauge-covariant derivative $D_\mu T = \nabla_\mu T + g_c \left[A_\mu,T\right]$ and field strength $F_{\mu\nu} = \partial_\mu A_\nu - \partial_\nu A_\mu + g_c\left[A_\mu,A_\nu\right]$. We set the dimensionful gauge coupling $g_c=1$ throughout, which fixes the AdS length.

When written in 2$\times$2 blocks, the background three-charge asymptotically AdS black brane solution \cite{Behrndt:1998jd} takes the form
\bea
T = T_0 &\equiv& \text{diag}(X_1 \id,X_2 \id ,X_3 \id) \label{eq:T0}
\\
A = A_0 dt &\equiv& \text{diag}(A_1 \sigma, A_2 \sigma, A_3 \sigma)dt \label{eq:A0}
\eea
where $\id$ is the 2$\times$2 unit matrix and $\sigma \equiv i \sigma_2$.  
The unimodularity condition on $T$ enforces $X_1X_2X_3=1$.  The black branes are parametrised by three charge parameters $q_I$ and take the form
\be
\begin{gathered}\label{gathered}
ds^2 = -H^{-2/3} f dt^2 + \frac{H^{1/3}}{f} dr^2 + r^2 H^{1/3} d \vec{x}_3^2 \\
X_I = \frac{H^{1/3}}{H_I}, \quad A_I =  \sqrt{\frac{r_+^4 H(r_+)}{q_I}} \left( \frac{1}{H_I(r)} - \frac{1}{H_I(r_+)} \right)
\end{gathered}
\ee
We have defined
\be
H_I(r) = 1 + \frac{q_I}{r^2}, \quad H(r) = H_1 H_2 H_3, \quad f(r) = - \frac{r_+^4 H(r_+)}{r^2} + r^2 H(r)
\ee
and $I=1,2,3$ labels each 2$\times$2 block. The horizon is at $r=r_+$; we use the coordinate freedom in $r$ to fix $r_+=1$ and thus $\kappa_I = q_I$ in the conventions of \cite{Son:2006em}. 

The thermodynamics of these backgrounds  was originally  studied in \cite{Chamblin:1999tk,Cvetic:1999ne} in the context of holography, with the study of spinning D3-branes, their counterparts in  the full 10D theory, initiated in \cite{Gubser:1998jb,Cai:1998ji} and extended in \cite{Cvetic:1999rb,Harmark:1999xt}.

\subsection{Multiple black brane branches}\label{sec:branches}
We work in the grand canonical ensemble throughout.
The chemical potentials for this system can be expressed in terms of the charge parameters
\be
\mu_I = \frac{\sqrt{q_I (1+q_1)(1+q_2)(1+q_3)}}{1+q_I},
\ee
and are conjugate to the three R-charges.   In this ensemble the relevant thermodynamic potential is the Gibbs potential, $\Omega$, with density $\omega =\frac{ \Omega}{\mathrm{vol}_3}$. We employ a dimensionless expression $\hat{\omega}$:
\be
\omega = -\frac{ (1+q_1)(1+q_2)(1+q_3)}{16 \pi G_5}, \qquad \hat{\omega} \equiv \frac{16\pi G_5\, \omega}{(\mu_1^2+\mu_2^2+\mu_3^2)^2}.
\ee
Similarly we employ a dimensionless version of the temperature, $\hat{T}$:
\be
T = \frac{2+q_1+q_2+q_3 -q_1 q_2 q_3 }{2\pi\sqrt{(1+q_1)(1+q_2)(1+q_3)}}, \qquad  \hat{T} = \frac{T}{\sqrt{\mu_1^2+\mu_2^2+\mu_3^2}}.
\ee

It is well known that this black brane family is not thermodynamically stable everywhere. For a stable thermodynamic equilibrium, the theory should lie at a local minimum of the thermodynamic potential, at fixed temperature and chemical potential. Constructing the Hessian matrix for variations of the energy density $\varepsilon$,
\be
\mathcal{H} = \frac{\partial^2 \varepsilon}{\partial s\partial \rho_I}
\ee
with respect to the entropy density $s$ and the R-charge densities $\rho_I$ following \cite{Son:2006em}, a necessary condition for local thermodynamic stability is given by
\be
\det\,\mathcal{H} \propto 2-q_1-q_2-q_3 + q_1 q_2 q_3 >0. \label{detHess}
\ee
However note that this is not a sufficient condition for local thermodynamic stability as there may be an even number of negative eigenvalues of $\mathcal{H}$.\footnote{We wish to thank Mukund Rangamani for bringing this point to our attention.}
At a fixed set of chemical potentials, $(\mu_1,\mu_2,\mu_3)$, we find that there is a minimum temperature for the existence of a thermodynamically stable dominant branch of solutions.

Without loss of generality we focus on  $\mu_I\geq0$ at which we only consider those solutions with $q_I\geq0$. This ensures that the solutions are regular. For this ensemble we fix two dimensionless ratios of the chemical potentials, $\zeta_1 = \frac{\mu_1}{\mu_3}$ and $\zeta_2 = \frac{\mu_2}{\mu_3}$, or any function thereof, in addition to $\hat{T}$. In general there are four black brane solutions at any grand canonical coordinate, $(\zeta_1,\zeta_2, \hat{T})$. To explore the phases of the system we adopt the coordinates
\be
p_1=\frac{\mu_1+\mu_2-2\mu_3}{\sqrt{6}\sqrt{\mu_1^2+\mu_2^2+\mu_3^2}},\qquad p_2= \frac{\mu_1-\mu_2}{\sqrt{2}\sqrt{\mu_1^2+\mu_2^2+\mu_3^2}}\label{pcoords}
\ee
which are dimensionless quantities and depend only on $\zeta_{1,2}$. 

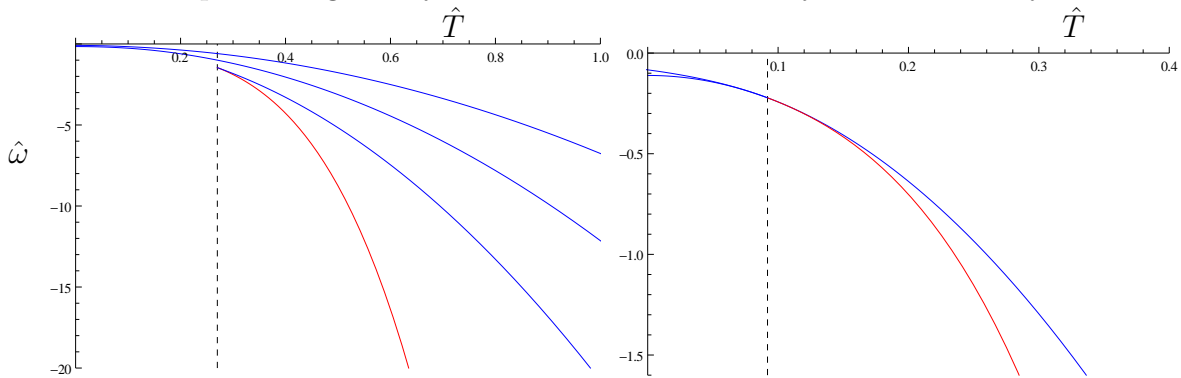
\begin{figure}[h!]
\begin{center}
\includegraphics[width=0.45\textwidth]{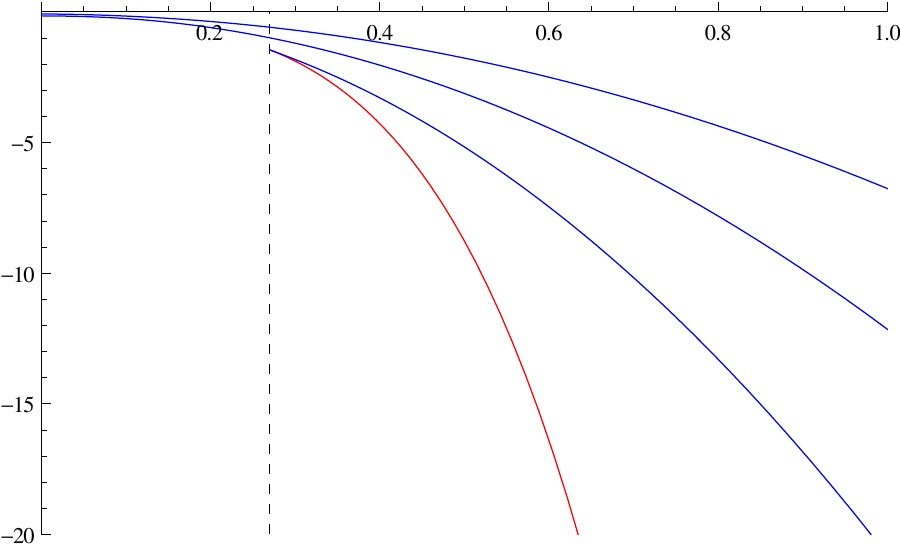}
\includegraphics[width=0.45\textwidth]{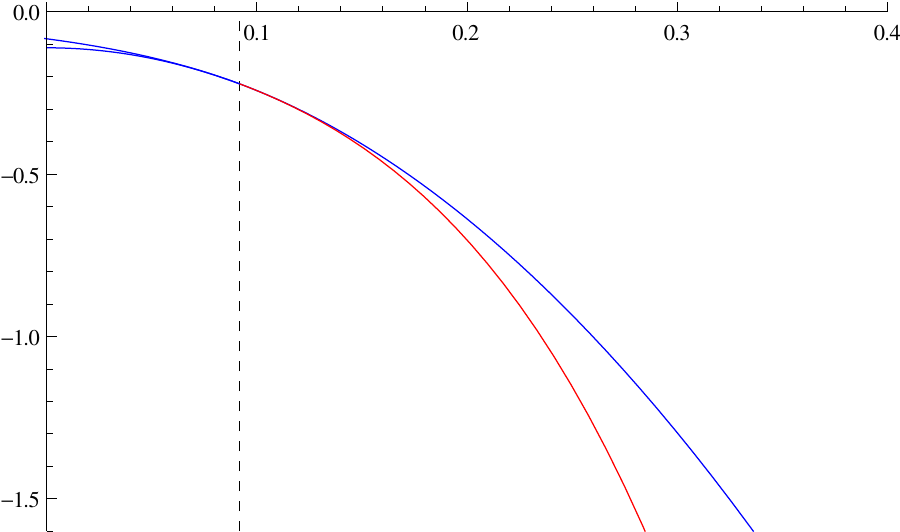}
\setlength{\unitlength}{0.1\columnwidth}
\begin{picture}(0.1,0.25)(0,0)
\put(-6.,2.85){\makebox(0,0){$\hat{T}$}}
\put(-1.,2.85){\makebox(0,0){$\hat{T}$}}
\put(-9.5,1.8){\makebox(0,0){$\hat{\omega}$}}
\end{picture}
\caption{\emph{Left:} Gibbs free energy density for the various black brane branches that exist at fixed $(p_1,p_2)=(-0.2,-0.1)$, defined in \eqref{pcoords}. Solutions that are thermodynamically unstable are shown in blue. A locally thermodynamically stable branch has a minimum temperature marked by the dashed line. \emph{Right:} Fixed $p_1=p_2=0$. The equal-charge black brane is only thermodynamically stable above the temperature indicated by the dashed line, whilst below it is both thermodynamically unstable and subdominant in the grand canonical ensemble.
}
\label{fig:freeenergy}
\end{center}
\end{figure}

At a representative fixed $(p_1,p_2)$ we show the free energy as a function of the temperature in the left panel of figure \ref{fig:freeenergy}, illustrating that the dominant line of black branes are only stable above a threshold temperature. This is the threshold for thermodynamic instability as described by the condition \eqref{detHess}, as well as the positivity of the eigenvalues of $\mathcal{H}$. In appendix  \ref{app:branch} we show that there is always one dominant branch of black branes with a minimum temperature given by the threshold for thermodynamic instability.

It is worth noting that in some cases there is a branch of black branes which satisfy the condition \eqref{detHess}, but which are not thermodynamically stable due to an even number of negative eigenvalues of $\mathcal{H}$. For example,  the three equal chemical potential case,  where $p_1=p_2=0$, is shown in the right panel of figure \ref{fig:freeenergy}.  The dominant black brane above the critical temperature $\hat{T}\simeq 0.09$ is locally thermodynamically stable and has equal charges $q_1=q_2=q_3$ and is nothing but the Reissner-Nordstr\"om-$AdS_5$ (RN) black brane. Below this critical temperature $\mathcal{H}$ picks up two equal negative eigenvalues with eigenvectors $\delta(s,\rho_1,\rho_2,\rho_3) = (0,-1,0,1),(0,-1,1,0)$. These eigenvectors for this thermodynamic instability are naturally associated with a Gubser-Mitra dynamical instability \cite{Gubser:2000ec, Gubser:2000mm}, as well as the generation of neutral zero modes at the threshold of stability which we discuss further in section \ref{sec:plusmodes}.

On a different tack, the free energy analysis presented here reveals a worrisome issue associated with the RN solution, as considered in this top-down approach. Working in the further consistent truncation of our theory corresponding to three equal charge parameters, RN becomes the only black brane solution and therefore appears to be thermodynamically dominant. However, below the critical temperature for the thermodynamic instability discussed above, $\hat{T}\simeq 0.09$, RN is not thermodynamically dominant in our more complete theory.

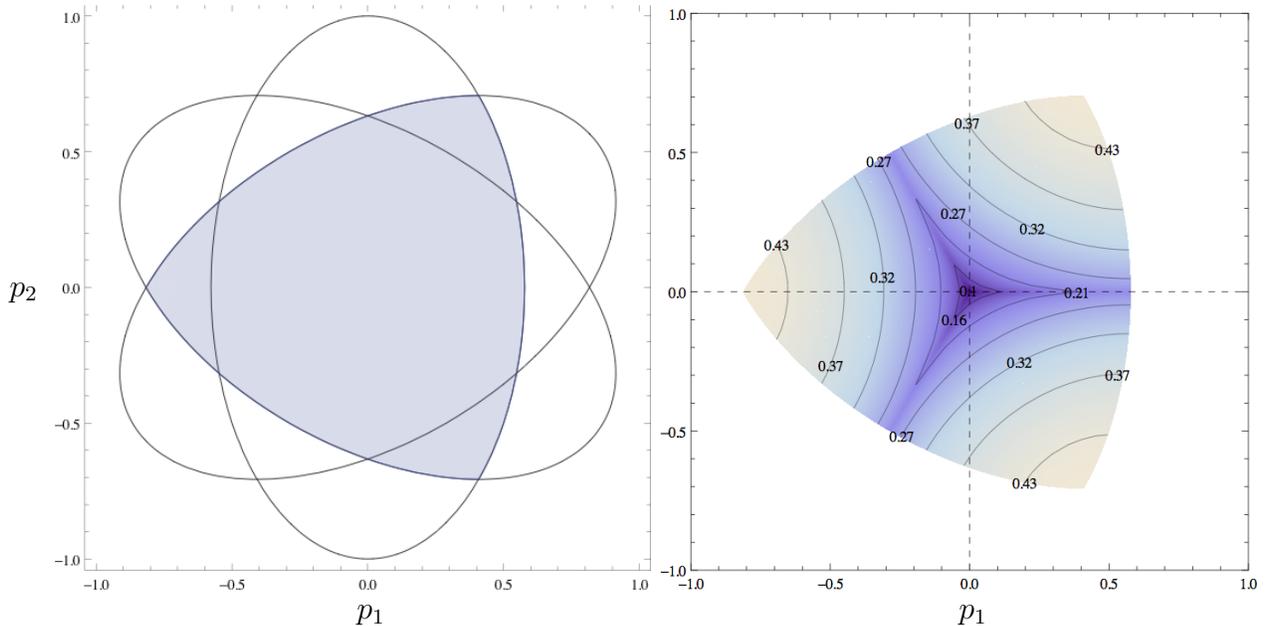
\begin{figure}[h!]
\begin{center}
\includegraphics[width=0.48\textwidth]{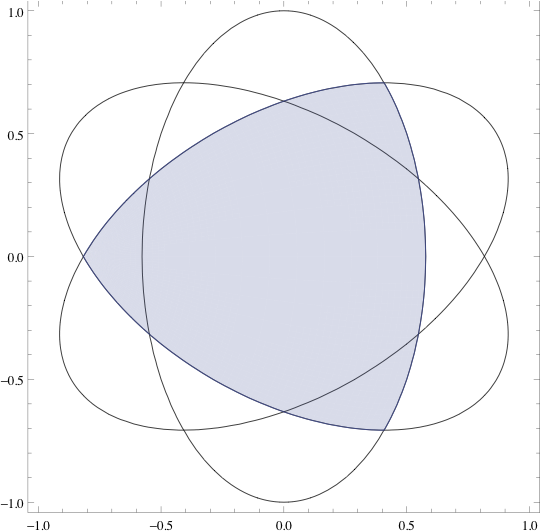}
\includegraphics[width=0.48\textwidth]{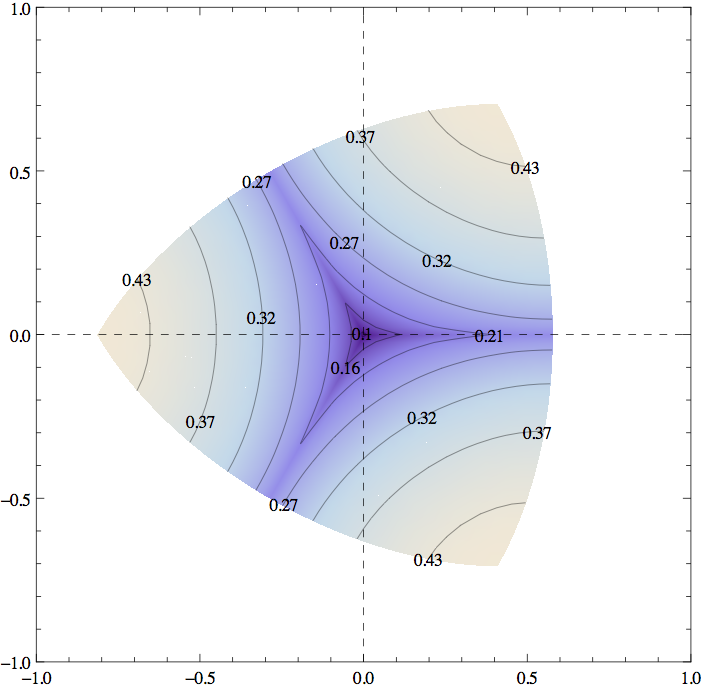}
\setlength{\unitlength}{0.1\columnwidth}
\begin{picture}(0.1,0.25)(0,0)
\put(-7.3,-0.2){\makebox(0,0){$p_1$}}
\put(-10.1,2.4){\makebox(0,0){$p_2$}}
\put(-2.45,-0.2){\makebox(0,0){$p_1$}}
\end{picture}
\caption{\emph{Left:} Region where black brane backgrounds with $\mu_I\geq0$ exist.  \emph{Right:} Critical temperatures for the thermodynamic instability --- \emph{i.e.}\ the minimum temperature, $\hat{T}$, for which the dominant branch is thermodynamically stable. \label{fig:background}}
\end{center}
\end{figure}

To understand the coordinates \eqref{pcoords} better, we project the condition $\mu_I\geq0$ into $(p_1,p_2)$ space. This region is formed by the intersection of three ellipses, with equations
\be
3 p_1^2 + p_2^2 = 1, \quad 3 p_1^2 + 5 p_2^2 \pm  2\sqrt{3} p_1 p_2
 = 2
\ee
and is shown in figure \ref{fig:background}. Within this region we find there is always a stable branch that dominates the ensemble above some critical temperature, as in the explicit examples shown in figure \ref{fig:freeenergy}. The minimum temperature for this dominant stable branch as a function of $p_1$ and $p_2$ is illustrated in the right panel of figure \ref{fig:background}. 

%

%%%%%%%%%%%%%%%%%%%%%%%%%%%%%%%%%%%%%%%%%%%%%%
\section{Fluctuations about the normal phase}\label{sec:fluctuations}
%%%%%%%%%%%%%%%%%%%%%%%%%%%%%%%%%%%%%%%%%%%%%%
We now detail the fluctuations of the 20 scalars and 15 gauge fields about the background normal phase solution given in section \ref{sec:trunc}.
We parameterise deviations from the scalar field matrix $T_0$ \eqref{eq:T0} by the fluctuation matrix $\chi(r)$, and deviations about the gauge field matrix $A_0$ \eqref{eq:A0} with radial electric fluctuations, $\alpha(r)$:
\be \label{eq:Tflucs}
T = T_0 e^{\chi}\, , \qquad \tr \chi = 0\, ; \qquad A = (A_0 + \alpha)dt
\ee
where the trace condition enforces unimodularity of $T$ at all orders in the fluctuation. The action for the linearised fluctuations can be found in appendix \ref{app:details}. Since the background is block diagonal, the fluctuations in different 2$\times $2 blocks do not couple. 

In general the fluctuations $\chi$ and $\alpha$ source metric fluctuations. For general background $\{q_I\}$ these may be switched off by enforcing the conditions
\be
\tr\, \chi_{II} =  \tr\, \sigma\alpha_{II} = 0 \label{diagonalconstraints}
\ee
where the trace is performed within the 2$\times$2 block. In detail these conditions are so restrictive that we have $\alpha_{II} = 0$ and $\sigma\chi_{II} + \chi_{II}\sigma = 0$, leaving the only non-zero diagonal block fluctuation as $ \sigma\chi_{II} - \chi_{II}\sigma$, which as we shall see shortly is a charged mode. For special cases of the background the conditions relax: for two or three equal charges we find
\be
\sum_I \tr\, \chi_{II} = \sum_I \tr\, \sigma\alpha_{II} = 0
\ee
where the sum is performed over equal $q_I$ blocks, with any remaining diagonal blocks constrained by \eqref{diagonalconstraints}.

\subsection{Diagonal fluctuations}
Without loss of generality we focus on fluctuations living in the $(1,1)$ block,
\be
\chi = \left(
\begin{array}{ccc}
\chi_{11} & 0 & 0 \\
0 & 0 & 0\\
0 & 0 & 0
\end{array}
\right),\qquad
\alpha = \left(
\begin{array}{ccc}
\sigma\hat{\alpha}_{11} & 0 & 0 \\
0 & 0 & 0\\
0 & 0 & 0
\end{array}
\right)\label{diagonal}
\ee
where $\chi_{11}$ is a symmetric 2$\times$2 matrix and $\hat{\alpha}_{11}$ is proportional to the 2$\times$2 identity matrix. To satisfy the backreaction and unimodularity constraints away from equal charges we must impose $\tr \chi_{11} = 0$. 

The equations of motion for these fluctuations may be partially diagonalised by moving to the new variables
\bea
\textbf{Neutral:}\quad\phi_{11}^{(+)} = \sigma \chi_{11} + \chi_{11} \sigma,&&\quad \alpha_{11}^{(+)} = \sigma \hat{\alpha}_{11} +\hat{\alpha}_{11} \sigma, \label{diagonalNewVars}\\
\textbf{Charged:}\quad\phi_{11}^{(-)} = \sigma \chi_{11} - \chi_{11} \sigma, &&\quad \alpha_{11}^{(-)} = \sigma \hat{\alpha}_{11} - \hat{\alpha}_{11} \sigma = 0.
\eea
The gauge potential for the $(-)$ mode vanishes due to the antisymmetry of $\sigma\hat{\alpha}_{11}$. Note that in what follows we have specialised to radial dependence in deriving these equations, despite their being presented in partially covariantised form. 

\subsubsection{Charged equations of motion}
The equation of motion for the charged scalar $\phi_{11}^{(-)}$ is
\be
\Box \phi_{11}^{(-)} -4 \left(X_1 (X_1-X_2-X_3)+ A_1^2\right)\phi_{11}^{(-)} = 0.\label{Weom}
\ee
Various special cases of this mode were considered nonlinearly in \cite{Aprile:2011uq} for the one, two and three equal-charge background; for these cases we have numerically verified the critical temperatures quoted. The unimodularity and backreaction constraint $\tr \chi_{11} = 0$ can be imposed without trivialising this mode.

\subsubsection{Neutral equations of motion} 
The equations of motion for the neutral scalar $\phi_{11}^{(+)}$ together with the gauge field fluctuation are
\bea
\Box \phi_{11}^{(+)} - 4X_1(X_1-X_2-X_3)\phi_{11}^{(+)}-\frac{F_1^2}{X_1^2}\phi_{11}^{(+)}+\frac{F_1\cdot h}{X_1^2}  = 0 \label{plusdiag1}\\
\frac{1}{\sqrt{-g}}\partial_r \left(\sqrt{-g}\;\frac{h^{rt}-2F_1^{rt}\phi_{11}^{(+)}}{X_1^2}\right)=0\label{plusdiag2}
\eea
where we have defined $h = d \alpha_{11}^{(+)}$.

If the mode is to be considered alone as in \eqref{diagonal}, we must impose $\tr \chi_{11} = 0$ for which we have $\phi_{11}^{(+)}=0$ by \eqref{diagonalNewVars}. In order to satisfy the backreaction and unimodularity constraints without trivialising these modes, we must therefore include other diagonal blocks, but this can only achieved if two or more of the background charges are equal. Such a sum of modes at two equal charges is closely related to the Gubser-Mitra instability \cite{Gubser:2000ec, Gubser:2000mm} --- a connection that we explore further in section \ref{sec:plusmodes}.

\subsection{Off-diagonal fluctuations}
Without loss of generality we focus on fluctuations living in the $(1,2)$ block (and $(2,1)$ by symmetry),
\be
\chi = \left(
\begin{array}{ccc}
0 & \chi_{12} & 0 \\
\chi_{12}^T & 0 & 0\\
0 & 0 & 0
\end{array}
\right),\qquad
\alpha = \left(
\begin{array}{ccc}
0 & \sigma\hat{\alpha}_{12} & 0 \\
\hat{\alpha}_{12}^T\sigma & 0 & 0\\
0 & 0 & 0
\end{array}
\right)\label{offdiagonal}
\ee
for which the condition $\tr \chi = 0$ and backreaction constraints are automatically satisfied.

As for the diagonal fluctuations, the equations of motion for these fluctuations may be partially diagonalised by moving to the new variables
\bea
\textbf{($+$) mode:}\quad\phi_{12}^{(+)} = \sigma \chi_{12} + \chi_{12} \sigma,&&\quad \alpha_{12}^{(+)} = \sigma \hat{\alpha}_{12} + \hat{\alpha}_{12} \sigma,\\
\textbf{($-$) mode:}\quad\phi_{12}^{(-)} = \sigma \chi_{12} - \chi_{12} \sigma, &&\quad \alpha_{12}^{(-)} = \sigma \hat{\alpha}_{12} - \hat{\alpha}_{12} \sigma.
\eea
where in contrast to the diagonal modes, here both the $(+)$ mode and $(-)$ mode are charged and couple to gauge field fluctuations in general. Dropping the indices and defining
\be
A_\pm \equiv \frac{1}{2}(A_1 \mp A_2),\qquad F_\pm \equiv \frac{F_1}{X_1}\pm\frac{F_2}{X_2},\qquad h_\pm \equiv d \alpha_\pm
\ee
we obtain the coupled equations of motion
\bea
\Box \phi_\pm  - \left(4 A_\pm^2 - 2 (X_1+X_2)X_3 +\frac{1}{4}F_\pm^2\right)\phi_\pm && \nonumber\\
+\frac{X_1+X_2}{ X_1 X_2}\left((X_1-X_2) A_\pm\cdot \alpha_
\pm  + \frac{1}{4}F_\pm\cdot h_\pm \right) & = & 0, \\
\frac{1}{2\sqrt{-g}}   \partial_r  \left[  \frac{\sqrt{-g}}{X_1X_2} \left(2h^{rt}_\pm -(X_1+X_2)  F^{rt}_\pm \phi_\pm \right) \right] && \nonumber\\
 - \frac{(X_1-X_2)^2}{X_1 X_2} g^{tt} \alpha_\pm + \frac{(X_1+X_2)(X_1-X_2)}{ X_1 X_2} A^t_\pm \phi_\pm & = & 0.
\eea

When $q_1 = q_2$  we have that $\phi_{12}^{(-)}$ satisfies the same equation of motion as the charged mode $\phi_{11}^{(-)}$ \eqref{Weom}, and $\phi_{12}^{(+)}$ satisfies the same equation of motion as the neutral mode $\phi_{11}^{(+)}$ \eqref{plusdiag1}. 

\subsection{Marginal modes}
Collecting all the modes above we have
\begin{center}
\begin{tabular}{c|c|c}
Type& Scalar field & See section\\
\hline
Charged diagonal & $\phi_{11}^{(-)},\phi_{22}^{(-)},\phi_{33}^{(-)}$  & \ref{sec:chargeddiag} \\
Off diagonal ($-$) & $\phi^{(-)}_{12},\phi^{(-)}_{13},\phi^{(-)}_{23}$  & \ref{sec:offdiagminus}\\
Neutral diagonal & $\phi_{11}^{(+)},\phi_{22}^{(+)},\phi_{33}^{(+)}$   & \ref{sec:plusmodes} \\
Off diagonal ($+$) & $\phi^{(+)}_{12},\phi^{(+)}_{13},\phi^{(+)}_{23}$ &  \ref{sec:plusmodes}\\
\end{tabular}
\end{center}
Note that the equation of motion for the $I\neq J$ mode $\phi_{IJ}^{(\pm)}$  reduces to the equation of motion for the $\phi_{II}^{(\pm)}$ when the background charges on the legs coincide: $q_I = q_J$. 

We seek time-independent solutions that are regular at the horizon and normalisable at the AdS boundary for each of the fluctuations detailed in the table above. Fluctuations of the gauge field at the boundary are required to vanish so that the chemical potentials are not perturbed. Such marginal modes indicate new branches of superconducting and normal black brane solutions. Developing series solutions we find the near-horizon behaviour (here the horizon position has been fixed to $r_+=1$)
\begin{align}
\phi &= a_1 + \ldots  +  \log(r-1) (a_2 + \ldots ) \\
\alpha &=a_3 (r-1)+ \ldots 
\end{align}
where the parameters $a_n$ are undetermined in the expansion and the ellipses denote determined higher order terms.  For regularity we set $a_2=0$, and  since the system of equations is  linear we are free to scale $a_1=1$. Additionally, for the diagonal modes we have fixed a shift symmetry in $\alpha$ such that $\alpha(r_+) = 0$. The scalar fluctuations in the UV vacuum all have mass $m^2 = -4$, which saturates the Breitenlohner-Freedman bound there.  The boundary series takes the form
\begin{align}
\phi &= r^{-2} (b_1 + \ldots ) +  r^{-2} \log r (b_2 + \ldots ) \\
\alpha &= b_3  + \ldots  +  r^{-2} (b_4 + \ldots )
\end{align}
with data $b_n$. Normalisability demands that we set $b_2=0$ and we must also impose $b_3=0$ to preserve our choice of chemical potentials.

%%%%%%%%%%%%%%%%%%%%%%%%%%%%%%%%%%%%%%%%%%%%%%
\section{Sector $(-)$}\label{sec:minusmodes}
%%%%%%%%%%%%%%%%%%%%%%%%%%%%%%%%%%%%%%%%%%%%%%
In this section we consider fluctuations of the fields labelled by $(-)$. 
Before considering general ratios of the chemical potentials and general $p_1,p_2$ coordinates it is instructive to study a slice of the allowed region (figure \ref{fig:background}) at $\mu_1=\mu_2$, for which $p_2=0$. We find marginal modes with critical temperatures illustrated in figure \ref{fig:slice}.

On this slice we can consider the effect of an imbalance on the mode $\phi_{33}^{(-)}$, which is charged under $A_3$ and does not couple directly to any other gauge field. At $p_2=0$, $p_1$ becomes a function of the imbalance for this mode, $\zeta_{1,2} = \frac{\mu_{1,2}}{\mu_3}$. The imbalance $\zeta_{1,2} = 0$ at $p_1 = -\sqrt{2/3}$, $\zeta_{1,2} = 1$ at $p_1=0$ and $\zeta_{1,2} \to \infty$ as $p_1 \to \sqrt{1/3}$. As shown in figure \ref{fig:slice} the critical temperature as a function of this imbalance meets the temperature of thermodynamic instability at sufficiently high $\zeta_{1,2}$. Note however that before this value of $\zeta_{1,2}$ is reached we see superconducting branches for the modes $\phi_{11}^{(-)},\phi_{22}^{(-)}$ and $\phi_{12}^{(-)}$, which are charged under $A_{1,2}$.

Extending the analysis to the full allowed region of $(p_1,p_2)$ space, we find the modes with the highest critical temperature as labelled in figure \ref{fig:highestTc}.

\begin{figure}[h!]
\begin{center}
\includegraphics[width=0.9\textwidth]{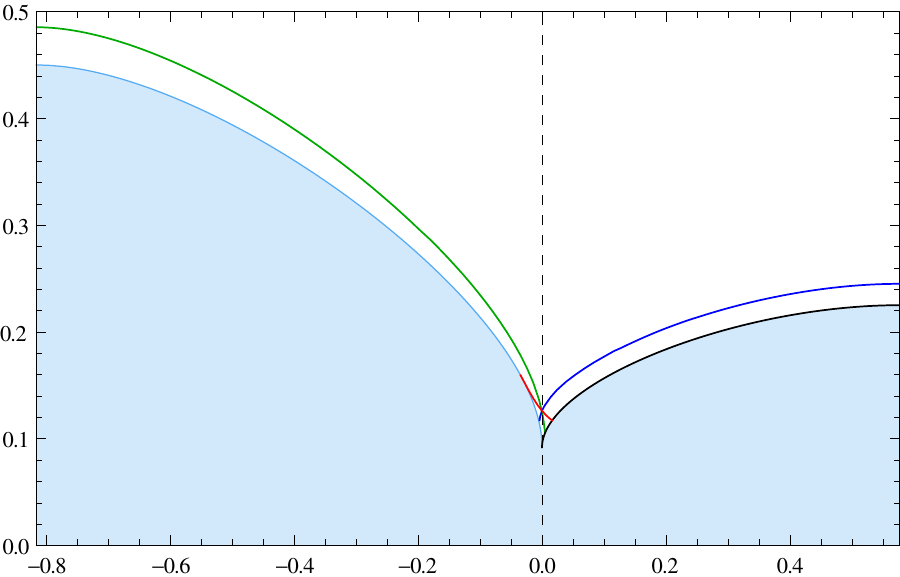}
\setlength{\unitlength}{0.1\columnwidth}
\begin{picture}(0.1,0.25)(0,0)
\put(-1.9,-0.1){\makebox(0,0){$p_1$}}
\put(-9.3,4.0){\makebox(0,0){$\hat{T}$}}
\end{picture}
\caption{A $\mu_1=\mu_2$ slice ($\mu_3\neq 0$) as a warm-up to the full $(p_1,p_2)$ space of normal phase solutions. The light blue region is thermodynamically unstable. The lines indicate critical temperatures of \emph{green:} $\phi_{33}^{(-)}$, \emph{red:} $\phi_{13}^{(-)},\phi_{23}^{(-)}$, \emph{blue:} $\phi_{11}^{(-)},\phi_{22}^{(-)},\phi_{12}^{(-)}$. Finally we show the line of neutral scalar zero modes (see section \ref{sec:plusmodes}) in \emph{black:} $\phi_{11}^{(+)},\phi_{22}^{(+)},\phi_{12}^{(+)}$ . \label{fig:slice}}
\end{center}
\end{figure}
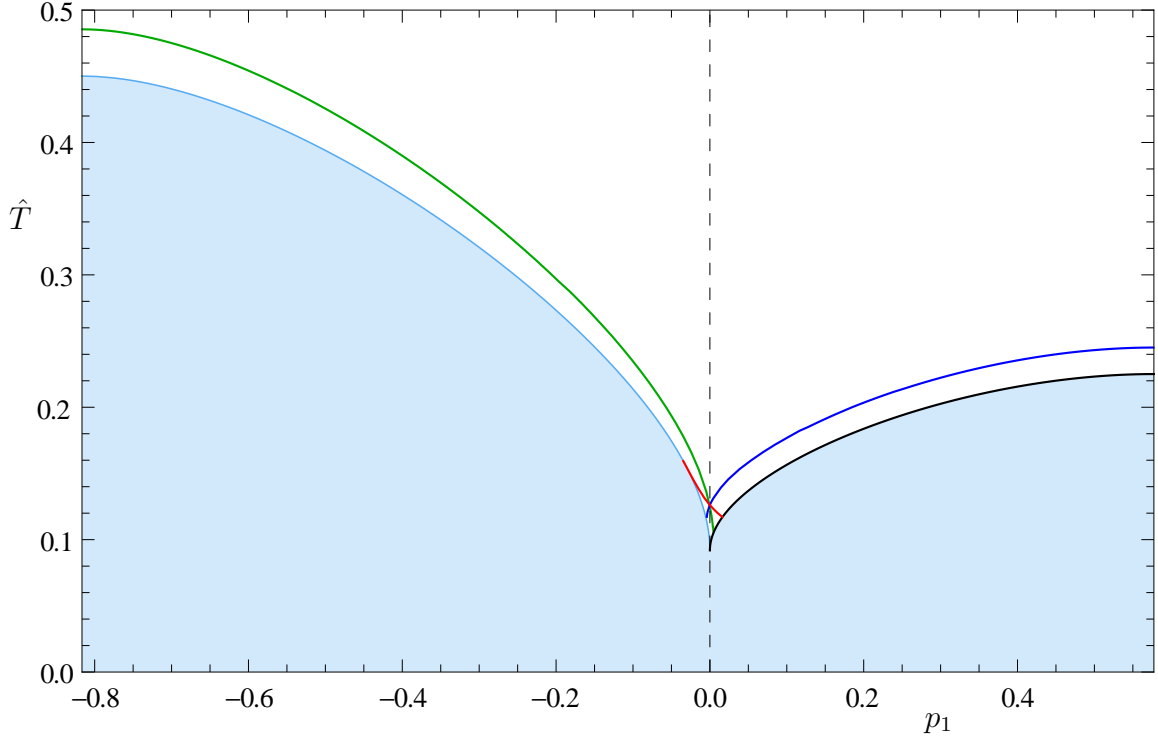
\begin{figure}[h!]
\begin{center}
\includegraphics[width=0.6\textwidth]{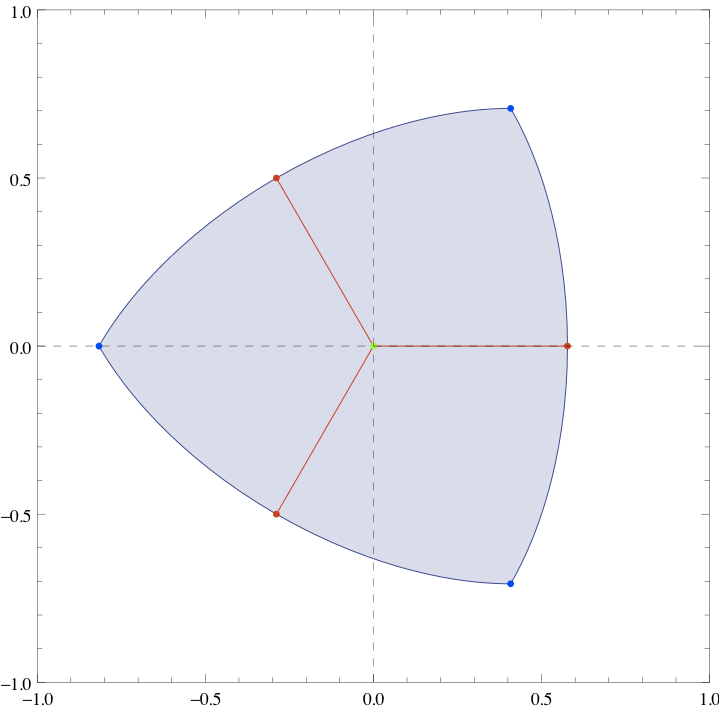}
\setlength{\unitlength}{0.1\columnwidth}
\begin{picture}(0.1,0.25)(0,0)
\put(-3.,-0.2){\makebox(0,0){$p_1$}}
\put(-6.6,3.0){\makebox(0,0){$p_2$}}
\put(-2.5,4.2){\makebox(0,0){$\phi^{(-)}_{11}$}}
\put(-2.5,2.0){\makebox(0,0){$\phi^{(-)}_{22}$}}
\put(-4.5,3.1){\makebox(0,0){$\phi^{(-)}_{33}$}}
\put(-2.2,3.1){\makebox(0,0){$\phi^{(-)}_{12}$}}
\put(-3.5,3.8){\makebox(0,0){$\phi^{(-)}_{13}$}}
\put(-3.5,2.2){\makebox(0,0){$\phi^{(-)}_{23}$}}
\put(-0.9,3.05){\makebox(0,0){$(\mu,\mu,0)$}}
\put(-4,4.7){\makebox(0,0){$(\mu,0,\mu)$}}
\put(-4,1.3){\makebox(0,0){$(0,\mu,\mu)$}}
\put(-5.8,3.2){\makebox(0,0){$(0,0,\mu)$}}
\put(-1.8,5.2){\makebox(0,0){$(\mu,0,0)$}}
\put(-1.8,0.8){\makebox(0,0){$(0,\mu,0)$}}
%\put(-3.1,3.05){\makebox(0,0){$(\mu,\mu,\mu)$}}
\end{picture}
\caption{Modes which give the highest critical temperature. The diagonal modes $\phi_{11}^{(-)},\phi_{22}^{(-)},\phi_{33}^{(-)}$ give the highest critical temperature in the blue regions as shown. In addition, along the red lines the modes  $\phi_{12}^{(-)},\phi_{13}^{(-)},\phi_{23}^{(-)}$ linearly coincide with the diagonal modes and have the same critical temperature. Dots indicate special cases of the chemical potentials, $(\mu_1,\mu_2,\mu_3)$, with all equal at the origin.\label{fig:highestTc}}
\end{center}
\end{figure}
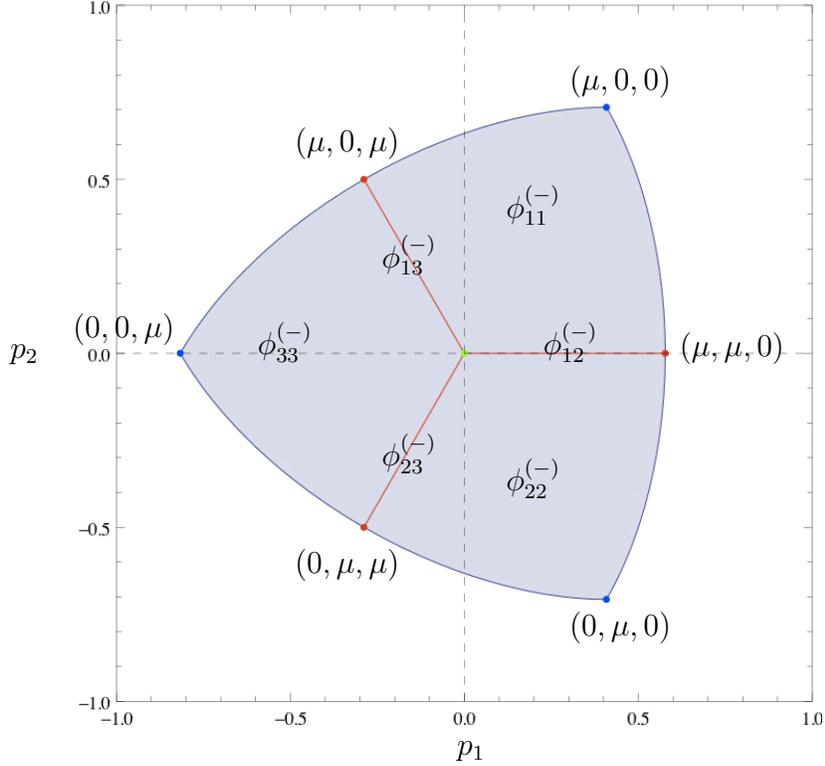

\subsection{$\phi^{(-)}_{11},\phi^{(-)}_{22},\phi^{(-)}_{33}$}\label{sec:chargeddiag}
\begin{figure}[h!]
\begin{center}
\includegraphics[width=0.6\textwidth]{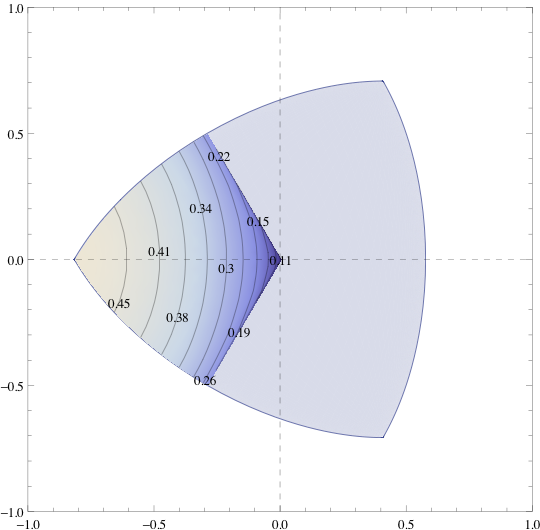}
\setlength{\unitlength}{0.1\columnwidth}
\begin{picture}(0.1,0.25)(0,0)
\put(-3.,-0.2){\makebox(0,0){$p_1$}}
\put(-6.6,3.0){\makebox(0,0){$p_2$}}
\end{picture}
\caption{Critical temperatures for the mode $\phi^{(-)}_{33}$. The modes $\phi^{(-)}_{11},\phi^{(-)}_{22}$ can be obtained by permuting the labels and appear predominantly in the regions as labelled in figure \ref{fig:highestTc}. \label{fig:chargeddiag}}
\end{center}
\end{figure}

The critical temperatures for $\phi^{(-)}_{33}$ are plotted in figure \ref{fig:chargeddiag}. It is important to emphasise that the region where  $\phi^{(-)}_{33}$ is unstable is not restricted to the segment indicated in figure \ref{fig:highestTc} --- there is a slight overlap of $\phi^{(-)}_{11},\phi^{(-)}_{22}$ and $\phi^{(-)}_{33}$ near lines of equal chemical potential.

\subsection{$\phi^{(-)}_{12},\phi^{(-)}_{13},\phi^{(-)}_{23}$}\label{sec:offdiagminus}
\begin{figure}[h!]
\begin{center}
\includegraphics[width=0.9\textwidth]{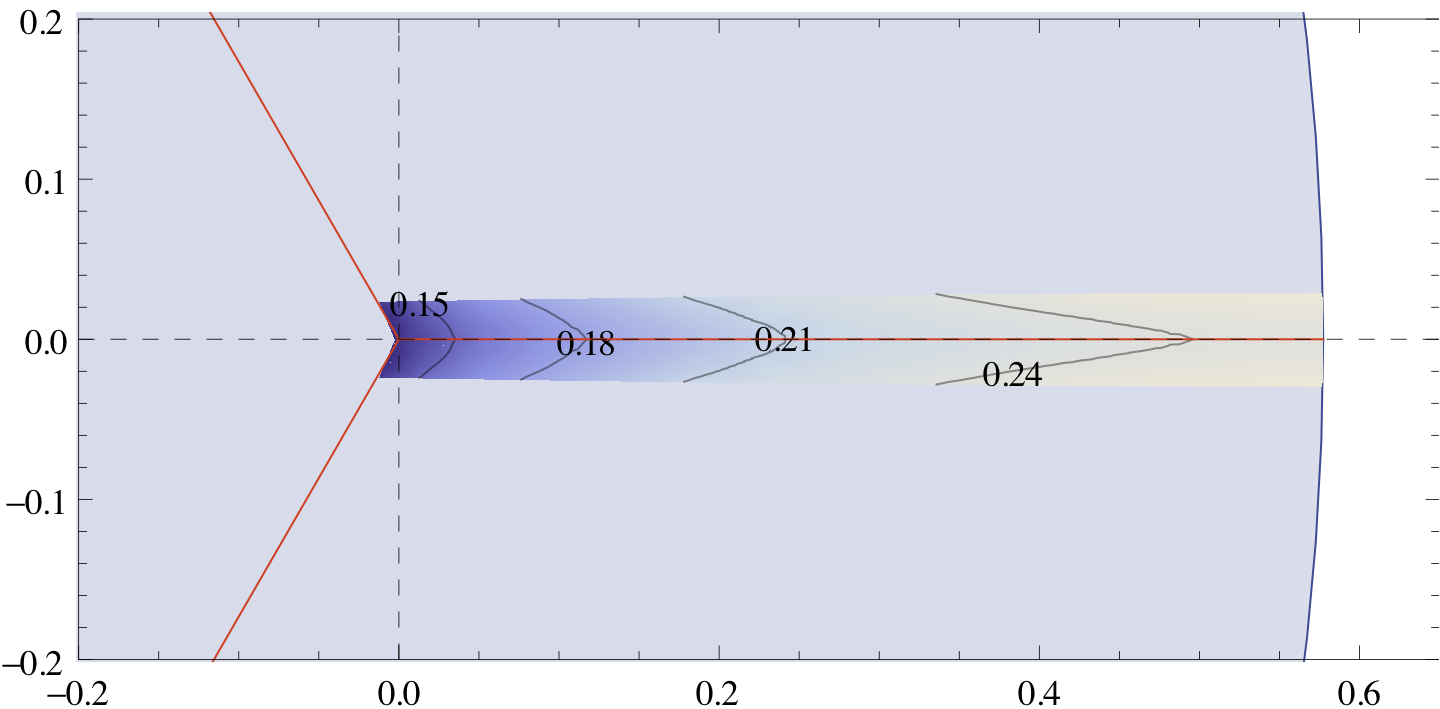}
\setlength{\unitlength}{0.1\columnwidth}
\begin{picture}(0.1,0.25)(0,0)
\put(-1.6,-0.05){\makebox(0,0){$p_1$}}
\put(-9.4,2.25){\makebox(0,0){$p_2$}}
\end{picture}
\caption{Region where marginal modes for $\phi^{(-)}_{12}$ exist, with critical temperatures indicated by the contours. Regions for the other modes $\phi^{(-)}_{23}$ and  $\phi^{(-)}_{13}$ can be found by permuting the indices. The blue shaded region is the region shown in figure \ref{fig:background}.  \label{fig:offdiagminus}}
\end{center}
\end{figure}

We find that marginal modes exist in a window of allowed $p_1$ and $p_2$ values.  This region and their critical temperatures are plotted in figure \ref{fig:offdiagminus}. Note that the critical temperatures all lie below those found in section \ref{sec:chargeddiag}, except along the locus of equal chemical potentials on the two legs; for instance along $p_2=0$ the critical temperatures for $\phi^{(-)}_{12}$ and $\phi^{(-)}_{11}$ and $\phi^{(-)}_{22}$ coincide.

\section{Sector $(+)$ and Gubser-Mitra instabilities}\label{sec:plusmodes}
In this section we consider fluctuations of the fields labelled by $(+)$. The fields $\phi_{11}^{(+)},\phi_{22}^{(+)},\phi_{33}^{(+)}$ are neutral, while $\phi^{(+)}_{12},\phi^{(+)}_{13},\phi^{(+)}_{23}$ become neutral when two chemical potentials coincide. We find marginal modes along the loci 
\bea
\phi_{11}^{(+)},\phi_{22}^{(+)},\phi^{(+)}_{12} &\quad \mu_1=\mu_2 >\mu_3\nonumber\\
\phi_{11}^{(+)},\phi_{33}^{(+)},\phi^{(+)}_{13} &\quad \mu_1=\mu_3 >\mu_2\\
\phi_{22}^{(+)},\phi_{33}^{(+)},\phi^{(+)}_{23} &\quad \mu_2=\mu_3 >\mu_1\nonumber
\eea
where all fluctuations are neutral. The backreaction and unimodularity constraints for the diagonal modes can be satisfied by turning on two modes, e.g. $\phi_{22}^{(+)} = -\phi_{11}^{(+)}$.

It is perhaps surprising that unlike the $(-)$ modes of section \ref{sec:minusmodes}, these marginal modes exist only for one dimensional regions of the $(p_1,p_2)$ plane.  However, they do not indicate a true instability of the black brane background.  Instead, as we shall demonstrate, they have a natural interpretation as zero modes which move in the family of background black branes. The existence of these zero modes is closely related to the threshold of thermodynamic instability; indeed, the critical temperatures are given precisely by the lowest temperature at which the normal phase is locally thermodynamically stable.  

We can analytically construct this class of marginal modes in the grand canonical ensemble by perturbing the charges and the horizon position whilst ensuring that the chemical potential and temperature remain fixed. A 4$\times$4 matrix for this variation is
\be
\mathcal{M} = \frac{\delta \{\mu_I, T\}}{\delta \{q_I,r_+\}} \label{muThess}
\ee
and, if the fluctuations are to be marginal modes in the grand canonical ensemble, we require $\det \mathcal{M} = 0$, which is achieved for background values of $q_I$ at the threshold of the thermodynamic instability given by $\det \mathcal{H}=0$. We emphasise that this applies to modes which include gravitational fluctuations at any values of the chemical potentials and is easily derived in this way. For example, when $q_3 = \frac{2-q_1-q_2}{1-q_1q_2}>0$ we have a surface of zero modes generated by perturbations to \eqref{gathered}, $H_I \to H_I + \delta H_I$, $f\to f+\delta f$, where
\bea
\delta H_i &=& \frac{q_i}{(q_i-1)}\frac{1}{r^2},\qquad\qquad i=1,2\\
\delta H_3 &=& -\frac{q_1+q_2-2}{(q_1-1)(q_2-1)}\frac{1}{r^2}\\
\delta f &=& 2 + \frac{ q_1 q_2 (q_1+q_2-2)}{(q_1q_2-1)}\left(\frac{1}{r^2}-\frac{1}{r^4}\right)
\eea
Specialising to the non-gravitational modes of interest here which exist only along loci of the chemical potentials, we obtain\footnote{The no-backreaction condition imposes that any two of the background charges are equal; let us here choose $q_1 = q_2$.  Subsequent conditions on the fluctuations themselves become $\delta q_1+\delta q_2 = 0$ and $\delta q_3=0$. Further imposing that the chemical potential is not perturbed we have $q_2 = 1$. }
\bea
\frac{\delta X_1}{X_1} = -\frac{\delta X_2}{X_2} &=& \frac{1}{1+r^2}, \qquad \delta X_3 = 0\\
\delta A_1 = -\delta A_2 &=&  \frac{\sqrt{(1+q_3)}(r^2-1)}{\left(1+r^2\right)^2}, \qquad \delta A_3 = 0
\eea
at temperature $\hat{T} = \frac{1}{\pi \sqrt{2(1+4q_3+q_3^2)}}$. This zero mode corresponds to a redistribution of the black brane charges when two of them are equal. These modes satisfy  equations \eqref{plusdiag1} and \eqref{plusdiag2} and consequently it is precisely this mode which gives rise to the black line in figure \ref{fig:slice}. It is also the zero mode generated by the thermodynamically unstable directions discussed in section \ref{sec:branches}.  We expect that there is also a true instability associated with these fluctuations \cite{Gubser:2000ec, Gubser:2000mm} but it is not captured by a zero mode analysis. We also anticipate a true instability away from the lines of equal chemical potentials, provided gravitational fluctuations are included.

%%%%%%%%%%%%%%%%%%%%%%%%%%%%%%%%%%%%%%%%%%%%%%
\section{Discussion}\label{sec:discussion}
%%%%%%%%%%%%%%%%%%%%%%%%%%%%%%%%%%%%%%%%%%%%%%
We have found new branches of superconducting black branes in $\mathcal{N}=8$ $SO(6)$ gauged supergravity, coming from a truncation involving 20 scalar fields and 15 gauge fields. Superconductivity persists to any values of the chemical potentials conjugate to the three R-charge densities of the background black brane.  The critical temperatures at which these black brane solutions arise is always higher than the temperature at which the three-charge background becomes thermodynamically unstable.

It is known that in some special cases --- such as two and three non-zero equal chemical potentials --- one has `retrograde condensation', \emph{i.e.}\ the branch of superconducting black branes is thermodynamically subdominant and the condensate exists only above a critical temperature \cite{Aprile:2011uq} (see also \cite{Buchel:2009ge, Donos:2011ut} in 4D).\footnote{For systems exhibiting retrograde condensation, the expectation is that superconducting ground states do not exist in the planar theory, but can exist when the boundary has spherical topology \cite{Gentle:2011kv}.} The other previously studied case of this system is that of one non-zero chemical potential, where the superconducting black brane is thermodynamically dominant \cite{Aprile:2011uq}. Here, in all known cases of retrograde condensation, we find that there is an additional marginal mode with the same critical temperature (see figure \ref{fig:highestTc}). We leave a nonlinear investigation of this branch, and whether it is dominant, to future work.

For a given mode we may interpret dialling the chemical potentials as introducing an imbalanced mixture as initiated in \cite{Erdmenger:2011hp, Bigazzi:2011ak}. The critical temperature surface cannot be extended to arbitrarily low temperatures as the imbalance is dialled, due the presence of a thermodynamic instability in the normal phase at finite temperature. At a given set of chemical potentials, different modes in the consistent truncation experience a different effective imbalance. Interestingly, we find that before a superconducting instability vanishes as a function of its imbalance, another appears to take its place.

The difference of the chemical potentials sets a scale for the system. It would be interesting to see if this scale sets a value of spatial momentum in this truncation resulting in an inhomogeneous, FFLO-like phase \cite{loff}.\footnote{For examples of spatially modulated phases in a truncation of IIB supergravity on $S^5$ see \cite{Donos:2011ff}.} It would also be interesting to construct all of the superconducting branches found nonlinearly, away from the critical temperature. This would allow the direct comparison of the free energy with the normal phase, and consequently the identification of the true phase boundary for the spontaneous breaking of a $U(1)$-bulk symmetry. 

For many of the linear modes studied in this paper, labelled by $(+)$, we have found only neutral marginal modes, whose role is to move one infinitesimally within the family of background black branes. These modes occur precisely at the threshold temperature for thermodynamic instability. As such the marginal modes do not generate an actual instability of the black brane. We have shown that it is possible to populate the entire minimum temperature surface (as a function of chemical potentials) with zero modes of the background black brane, which can be found analytically and involve gravitational fluctuations in general. One expects to find a true instability associated with the thermodynamic instability in these cases.

Our work was motivated in part by the search for universal features in top-down holographic superconductivity. We wish to highlight a basic problem we have encountered when using consistent truncations: a subdominant solution in a given theory can appear as the dominant solution in a consistent truncation of that theory. This is exemplified by the normal phase solutions employed in this paper.

In the case of three equal chemical potentials the normal phase solution is simply RN at high temperatures. Whilst the RN branch exists at low temperatures too, it is no longer the dominant contribution to the ensemble. This has important consequences for the equal-$q_I$ consistent truncation where RN is the only normal phase solution remaining. Evidently in the equal-$q_I$ truncation RN appears as the dominant black brane branch at all temperatures, when in fact, at low temperatures in the full theory it is subdominant to a thermodynamically unstable black brane.

\subsection*{Acknowledgements}
It is a pleasure to thank Aristomenis Donos, Jerome Gauntlett, Mukund Rangamani and Julian Sonner for helpful discussions. SAG is supported by a STFC studentship. BW is supported by the Royal Commission for the Exhibition of 1851.

\appendix
\section{Dominant branch}\label{app:branch}
Using $q_3$ as a parameter along the branch of solutions and writing $s_i^2 = (1+q_3)^2 - 4 q_3 \zeta_i^2$ we find that $q_1,q_2$ are given by
\be
q_1 = \frac{1+q_3-s_1}{1+q_3+s_1}, \qquad q_2 = \frac{1+q_3-s_2}{1+q_3+s_2}.
\ee
There are therefore up to four solutions coming from the sign choices for $s_1$ and $s_2$. In these coordinates the thermodynamic potential density becomes
\be
%\hat{T} = \frac{2+s_1+s_2}{4\pi \sqrt{q_3(1+\zeta_1^2+\zeta_2^2)}},\qquad  
\hat{\omega} = -\frac{(1+q_3)(1+q_3+s_1)(1+q_3+s_2)}{4 q_3^2(1+\zeta_1^2+\zeta_2^2)^2}
\ee
and so we see that at some desired grand canonical coordinate, $(\zeta_1,\zeta_2, \hat{T})$, the dominant configurations live on the $s_1,s_2>0$ branch, provided it exists there. The temperature is stationary at the threshold for thermodynamic stability:
\bea
\frac{\partial \hat{T}}{\partial q_3}\Bigg|_{\zeta_1,\zeta_2} & = & -\frac{(1-q_3^2) (s_1+s_2) + 2 s_1 s_2}{8\pi q_3 s_1 s_2  \sqrt{q_3(1+\zeta_1^2+\zeta_2^2)}}\\
\det\,\mathcal{H} & \propto & 2\frac{(1-q_3^2)(s_1+s_2) + 2 s_1 s_2}{(1+q_3+s_1)(1+q_3+s_2)}
\eea
and furthermore, on the $s_1,s_2>0$ branch the temperature is a minimum there: $\frac{\partial^2 \hat{T}}{\partial q_3^2}>0$. This shows that there is a branch of black branes which only exists above the threshold temperature for thermodynamic stability, where it dominates the ensemble.
\section{Action for fluctuations}\label{app:details}

In this appendix we present the action to quadratic order for fluctuations of \eqref{eq:action} of the form \eqref{eq:Tflucs} that do not source metric fluctuations:
\be
S_2 = \frac{1}{16\pi G_5} \int d^5x\sqrt{-g}\, \tr \left(L_\chi + L_\alpha + L_\text{int}\right)
\ee
with
\bea
L_\chi &=& -\frac{g^{rr}}{4}\chi'^2 - \frac{g_c^2g^{tt}}{2} \left((A_0 \chi)^2 - A_0^2 \chi^2\right) + \frac{g^{rr}g^{tt}}{4}\left( (T_0^{-1} A_0')^2 \chi^2+ (T_0^{-1} A_0'\chi)^2\right)\\
&&-g_c^2\left(T_0^2 -\frac{1}{2} (\tr T_0) T_0\right)\chi^2 - g_c^2 (T_0 \chi)^2 \nonumber \\
L_\alpha &=& -\frac{g_c^2g^{tt}}{2}\left(\alpha^2 - T_0^{-1}\alpha T_0\alpha\right)+\frac{g^{rr}g^{tt}}{4} (T_0^{-1} \alpha' )^2\\
L_\text{int} &=& -\frac{g_c^2g^{tt}}{2} T_0^{-1} \alpha T_0 \left(A_0 \chi - \chi A_0\right) - \frac{g^{rr}g^{tt}}{2}\left(\chi T_0^{-1} A_0' + T_0^{-1} A_0' \chi\right) T_0^{-1} \alpha'
\eea
%\bibliography{STU}
%\bibliographystyle{utphys}

\providecommand{\href}[2]{#2}\begingroup\raggedright\endgroup

\end{document}